\documentclass[12pt]{article}
\usepackage{fancyhdr}
\usepackage{mathrsfs}
\usepackage[T1]{fontenc}
\usepackage{mathpazo}
\usepackage{setspace}
\usepackage{amsfonts}
\usepackage{amssymb}
\usepackage{amsmath}
\usepackage{epsfig}
\usepackage{latexsym}
\usepackage{color}
\usepackage{graphicx}
\usepackage{nicefrac}
\usepackage[latin1]{inputenc}
\usepackage{slashed}
\usepackage{multirow}
\usepackage{comment}
\usepackage{soul}
\usepackage{hyperref}
\usepackage[nosort]{cite}
\usepackage{datetime}

\usepackage[titletoc]{appendix}

\def\hybrid{\topmargin -20pt    \oddsidemargin 0pt
        \headheight 0pt \headsep 0pt
        \textwidth 6.25in       
        \textheight 9.25in       
        \marginparwidth .875in
        \parskip 5pt plus 1pt   \jot = 1.5ex}

\hybrid

\def\baselinestretch{1.2}

\catcode`\@=11

\def\marginnote#1{}
\newcount\hour
\newcount\minute
\newtoks\amorpm
\hour=\time\divide\hour by60
\minute=\time{\multiply\hour by60 \global\advance\minute by-\hour}
\edef\standardtime{{\ifnum\hour<12 \global\amorpm={am}%
        \else\global\amorpm={pm}\advance\hour by-12 \fi
        \ifnum\hour=0 \hour=12 \fi
        \number\hour:\ifnum\minute<10 0\fi\number\minute\the\amorpm}}
\edef\militarytime{\number\hour:\ifnum\minute<10 0\fi\number\minute}

\def\draftlabel#1{{\@bsphack\if@filesw {\let\thepage\relax
   \xdef\@gtempa{\write\@auxout{\string
      \newlabel{#1}{{\@currentlabel}{\thepage}}}}}\@gtempa
   \if@nobreak \ifvmode\nobreak\fi\fi\fi\@esphack}
        \gdef\@eqnlabel{#1}}
\def\@eqnlabel{}
\def\@vacuum{}
\def\draftmarginnote#1{\marginpar{\raggedright\scriptsize\tt#1}}

\def\draft{\oddsidemargin -.5truein
        \def\@oddfoot{\sl preliminary draft \hfil
        \rm\thepage\hfil\sl\today\quad\militarytime}
        \let\@evenfoot\@oddfoot \overfullrule 3pt
        \let\label=\draftlabel
        \let\marginnote=\draftmarginnote
   \def\@eqnnum{(\theequation)\rlap{\kern\marginparsep\tt\@eqnlabel}%
\global\let\@eqnlabel\@vacuum}  }

\def\preprint{\twocolumn\sloppy\flushbottom\parindent 2em
        \leftmargini 2em\leftmarginv .5em\leftmarginvi .5em
        \oddsidemargin -.5in    \evensidemargin -.5in
        \columnsep .4in \footheight 0pt
        \textwidth 10.in        \topmargin  -.4in
        \headheight 12pt \topskip .4in
        \textheight 6.9in \footskip 0pt
        \def\@oddhead{\thepage\hfil\addtocounter{page}{1}\thepage}
        \let\@evenhead\@oddhead \def\@oddfoot{} \def\@evenfoot{} }

\def\numberbysection{\@addtoreset{equation}{section}
        \def\theequation{\thesection.\arabic{equation}}}

\def\underline#1{\relax\ifmmode\@@underline#1\else
        $\@@underline{\hbox{#1}}$\relax\fi}

\def\titlepage{\@restonecolfalse\if@twocolumn\@restonecoltrue\onecolumn
     \else \newpage \fi \thispagestyle{empty}\c@page\z@
        \def\thefootnote{\fnsymbol{footnote}} }

\def\endtitlepage{\if@restonecol\twocolumn \else \newpage \fi
        \def\thefootnote{\arabic{footnote}}
        \setcounter{footnote}{0}}  

\catcode`@=12
\relax

\def\figcap{\section*{Figure Captions\markboth
        {FIGURECAPTIONS}{FIGURECAPTIONS}}\list
        {Figure \arabic{enumi}:\hfill}{\settowidth\labelwidth{Figure
999:}
        \leftmargin\labelwidth
        \advance\leftmargin\labelsep\usecounter{enumi}}}
 \relax
\def\tablecap{\section*{Table Captions\markboth
        {TABLECAPTIONS}{TABLECAPTIONS}}\list
        {Table \arabic{enumi}:\hfill}{\settowidth\labelwidth{Table
999:}
        \leftmargin\labelwidth
        \advance\leftmargin\labelsep\usecounter{enumi}}}
 \relax
\def\reflist{\section*{References\markboth
        {REFLIST}{REFLIST}}\list
        {[\arabic{enumi}]\hfill}{\settowidth\labelwidth{[999]}
        \leftmargin\labelwidth
        \advance\leftmargin\labelsep\usecounter{enumi}}}
 \relax
\makeatletter
\newcounter{pubctr}
\def\publist{\@ifnextchar[{\@publist}{\@@publist}}
\def\@publist[#1]{\list
        {[\arabic{pubctr}]\hfill}{\settowidth\labelwidth{[999]}
        \leftmargin\labelwidth
        \advance\leftmargin\labelsep
        \@nmbrlisttrue\def\@listctr{pubctr}
        \setcounter{pubctr}{#1}\addtocounter{pubctr}{-1}}}
\def\@@publist{\list
        {[\arabic{pubctr}]\hfill}{\settowidth\labelwidth{[999]}
        \leftmargin\labelwidth
        \advance\leftmargin\labelsep
        \@nmbrlisttrue\def\@listctr{pubctr}}}
 \relax
\makeatother
\newskip\humongous \humongous=0pt plus 1000pt minus 1000pt

\newif\ifdtup

\relax

\def\be{\begin{equation}}
\def\ee{\end{equation}}
\def\ba{\begin{eqnarray}}
\def\ea{\end{eqnarray}}

\def\del{\partial}

\def\a{\alpha}

\def\b{\beta}

\def\g{\gamma}

\def\d{\delta}

\def\p{\pi}

\def\m{\mu}
\def\n{\nu}

\def\l{\lambda}
\def\L{\Lambda}
\def\s{\sigma}

\def\cL{{\cal L}}

\def\no{\noindent}

\def\qq{\qquad}

\def\IR{\relax{\rm I\kern-.18em R}}

\def \ha {{1\over 2}}

\def \ov {\over}

\def\IR{\relax{\rm I\kern-.18em R}}
\def\IL{\relax{\rm I\kern-.18em L}}

\def\inv{^{\raise.15ex\hbox{${\scriptscriptstyle -}$}\kern-.05em 1}}

\def\cL{{\cal L}}

\def\tr{{\rm tr}}
\def\Tr{{\rm Tr}}

\begin{document}

\renewcommand{\theequation}{\thesection.\arabic{equation}}
\csname @addtoreset\endcsname{equation}{section}

\newcommand{\beq}{\begin{equation}}
\newcommand{\eeq}[1]{\label{#1}\end{equation}}
\newcommand{\ber}{\begin{equation}}
\newcommand{\eer}[1]{\label{#1}\end{equation}}
\newcommand{\eqn}[1]{(\ref{#1})}
\begin{titlepage}
\begin{center}


${}$
\vskip .2 in
\vskip .2 in
\vskip .2 in
\vskip .2 in

{\Large\bf 

Hamiltonian integrability of the webs of integrable theories
}

\vskip 0.4in

{\large\bf George Georgiou}\ \ 
\vskip 0.15in

 {\em
Department of Nuclear and Particle Physics,\\
Faculty of Physics, National and Kapodistrian University of Athens,\\
Athens 15784, Greece\\
}
\vskip 0.12in

{\footnotesize \texttt george.georgiou@phys.uoa.gr}


\vskip .5in
\end{center}

\centerline{\bf Abstract}

\no
We present the Hamiltonian formulation of the recently constructed integrable theories of  \cite{Georgiou:2020wwg}. These theories turn out to be canonically equivalent to the sum of an asymmetrically gauged CFT and of the most general  $\lambda$-deformed model of \cite{Georgiou:2018gpe}. Using the Hamiltonian formalism, we prove that the full set of conserved charges of the models of  \cite{Georgiou:2020wwg} are in involution, ensuring their Hamiltonian integrability. Finally, we show that the equations of motion of these theories can be put in the form of zero curvature Lax connections.

\vskip .4in
\noindent
\end{titlepage}
\vfill
\eject

\newpage

\tableofcontents

\noindent

\def\baselinestretch{1.2}
\baselineskip 20 pt
\noindent


\setcounter{equation}{0}

\section{Introduction}

Integrable structures are ubiquitous in theoretical physics. They appear in the context of gauge/gravity dualities, in the high energy regime of QCD, in the description of certain condensed matter systems and primarily in the context of 2-dimensional field theories.
They belong to the quiver of the most useful tools in  obtaining exact results in quantum field theory.

Integrable non-linear $\s$-models have played  an instrumental role in the context of gauge/gravity dualities. Thanks to the duality, one can map the strongly coupled dynamics of gauge theory  to the 
weakly coupled dynamics of an integrable two-dimensional non-linear $\s$-model. The prototypical example of such an integrable  $\s$-model is the principal chiral model (PCM) with or without  a Wess-Zumino (WZ) term and based on a semi-simple group $G$. Subsequently, it was realised that PCMs based on a semi-simple groups $G$ admit an integrable deformation that depends on an additional  continuous parameter \cite{Klimcik:2002zj,Klimcik:2008eq,Klimcik:2014}. These deformed integrable models are called Yang-Baxter (YB) models.  Furthermore,  YB models on  symmetric and semi-symmetric spaces were studied in  \cite{Delduc:2013fga,Delduc:2013qra,Arutyunov:2013ega}.
There  also exist two parameter integrable deformations of the PCM, namely the YB $\s$-model with a WZWN term \cite{Delduc:2014uaa} and the bi-YB model \cite{Klimcik:2014}. Moreover, integrable deformations of the PCM with three or more parameters were the subject of study  in \cite{Delduc:2017fib}.
It is a remarkable fact that all these models lie under the unifying description of the so-called $\cal E$-models \cite{Klimcik:2015gba,Klimcik:2017ken}. 
More recently, the systematic and powerful construction of classically integrable theories that is based on their interpretation as explicit realisations of affine Gaudin models \cite{Vicedo:2017cge} having an arbitrary number of sites was presented in \cite{Delduc:2019bcl,Delduc:2018hty} (see also \cite{Bassi:2019aaf}). The one and two loop renormalisation group (RG) flows of some of these models were derived in \cite{Delduc:2020vxy} and \cite{Levine:2021fof}, respectively.

In a parallel development, the systematic construction of the, all-loop in the deformation parameters, effective actions of a large class of integrable two-dimensional field theories based on group, symmetric and semi-symmetric spaces  was deployed in a series of papers \cite{Sfetsos:2013wia,Georgiou:2016urf,Georgiou:2017jfi,Georgiou:2018hpd,Georgiou:2018gpe,Hollowood:2014rla,Hollowood:2014qma,Driezen:2019ykp}. 
The aforementioned  models  contain several couplings. For small values of these couplings the models take the form of  one or more WZW models \cite{Witten:1983ar,Witten:1991mm}  perturbed by marginally relevant operators that are bi-linear in the currents. 
Following their construction, the quantum properties of these theories were studied in great detail 
\cite{Georgiou:2015nka,Georgiou:2016iom,Georgiou:2016zyo,Georgiou:2017oly,Itsios:2014lca,Georgiou:2017aei}. 
Many observables of these theories, such as their $\beta$-functions  \cite{Kutasov:1989dt,Gerganov:2000mt,Itsios:2014lca,Sfetsos:2014jfa,LeClair:2001yp,Appadu:2015nfa}, the anomalous dimensions of currents and primary operators \cite{Georgiou:2015nka,Georgiou:2016iom,Georgiou:2016zyo,Georgiou:2019jcf,Georgiou:2019aon} and three-point correlation functions  involving  currents and/or primary fields \cite{Georgiou:2015nka,Georgiou:2019aon,Georgiou:2020bpx} were computed as {\it exact} functions of the deformation parameters.
Moreover, the Zamolodchikov's C-functions \cite{Zamolodchikov:1986gt} of these models were calculated also
as {\it exact} functions of the deformation parameters \cite{Georgiou:2018vbb,Sagkrioti:2018abh}.\footnote{These results although exact in the deformation parameters give only the leading contribution in the $1/k$-expansion. Recently, the sub-leading terms in $1/k$-expansion for the $\beta$-functions were obtained in \cite{Georgiou:2019nbz,Hoare:2019mcc,Hassler:2020wnp}. 
For the C-function and the anomalous dimensions of the operators perturbing the CFT  the sub-leading  in $1/k$-expansion terms were calculated in  \cite{Georgiou:2019nbz} for the cases of group and coset spaces.}

Compared to the prototype single $\l$-deformed model of \cite{Sfetsos:2013wia} (for the group $SU(2)$ the $\l$-deformed model was found earlier in \cite{Balog:1993es}) most of the models mentioned above have RG flows that exhibit a rich structure consisting of several fixed points, with  different CFTs
sitting at different fixed points. The complete classification of these CFTs according to their symmetry groups remains an open problem. 
This goal was achieved in \cite{Georgiou:2020eoo} for a generalisation of the cyclic $\l$-deformed models of \cite{Georgiou:2017oly} in which  different levels for the WZW models were allowed.

\no

Even more recently, a class of integrable models was introduced in \cite{Georgiou:2020wwg}. These models couple $N$ WZW models at arbitrary levels with $n\le N$  of the following integrable theories, namely $n_1$  copies of  the PCM, $n_2$ copies of the YB model, both based on a group $G$,   $n_3$  copies of the   the isotropic $\s$-model on the symmetric coset space $G/H$ and $n_4$ copies of the YB model on the symmetric space $G/H$, with $n=n_1+n_2+n_3+n_4$.\footnote{As mentioned in \cite{Georgiou:2020wwg},  a suitable redefinition may eliminate  one of the deformation parameters in the case of the YB model on the symmetric space $G/H$ for all the examples worked out so far. Consequently, it remains to be seen if there exist any non-trivial examples for this type of deformation.}

The author  of \cite{Georgiou:2020wwg} argued integrability in  the following way. $n$ distinct combinations of the equations of motion were equivalent to zero curvature conditions of certain Lax connections.
The remaining $N-n$ equations of motion were shown
to take the form of covariantly free combinations of currents. Each of these covariantly free equations produces an infinite tower of local conserved  charges which supplement the ones obtained from the monodromy matrix constructed from the spatial component of the Lax pairs. As a result, one gets as many infinite towers of conserved 
charges as the degrees of freedom of the theories which indicates that the theories are classically integrable.
However, to complete the proof of integrability, one should also prove that all the conserved charged mentioned above are in involution.

The aim of the present paper is twofold. Firstly, to show that all the charges are, indeed, in involution. And secondly, to clarify the relation 
of the integrable models in \cite{Georgiou:2020wwg} to the integrable models that already exist in the literature.\footnote{In particular, we will see that the models constructed in \cite{Georgiou:2020wwg} are canonically equivalent to the sum of integrable models already present in the literature, namely of the gauged WZW model and of the most general $\l$-deformed model of \cite{Georgiou:2018gpe}. The precise relation of the integrable theories in \cite{Georgiou:2020wwg} to the affine Gaudin models \cite{Delduc:2019bcl} or the 4-d Chern-Simons formulation of integrable theories \cite{Costello:2019tri} remains to be clarified.} Furthermore, we will also show that our models are integrable in the more conventional way, namely that it is the {\it full set of equations of motion } that can be recast in the form of zero curvature conditions of certain Lax connections.

The plan of the paper is as follows. In section \ref{Rev}, we will very briefly review the models of \cite{Georgiou:2020wwg}. For more details one may refer to  \cite{Georgiou:2020wwg}. In section \ref{2}, we derive the Hamiltonians corresponding to the three integrable cases which are in turn considered in sections \ref {2.1.1},  \ref {2.1.2} and \ref {2.2}. Then, we prove that the whole set of conserved charges found in \cite{Georgiou:2020wwg} are in involution. 
The Hamiltonian of the theory is canonically equivalent to the sum of of the Hamiltonian of an asymmetrically gauged CFT plus the Hamiltonian of the most general  $\lambda$-deformed model of \cite{Georgiou:2018gpe} in the case where
level conservation at each vertex  is imposed.\footnote{For the models considered in section \ref{2.1} the matrix $(\l^{-1})^{ab}_{ij}$ of the most general $\l$-deformed model takes certain special values, namely $(\l^{-1})^{ab}_{ij}=\d_{ij}(\l^{-1}_i)^{ab}$.} 
In section \ref{3}, we show that the full set of equations of motion of our models can be rewritten as zero curvature conditions of certain Lax connections. Finally, in section \ref{4}, we present our conclusions.

\section{Review of the webs of integrable theories}\label{Rev}

In what follows, we shall briefly review the construction of the models presented in  \cite{Georgiou:2020wwg}.
These models couple  N WZW models  with an arbitrary number $n\leq N$ of the following four fundamental integrable theories, namely the PCM,  the YB model, both based on a group $G$,  the isotropic $\s$-model on the symmetric coset space $G/H$ and  the YB model on the symmetric space $G/H$. The coupling is achieved by gauging 
the left global symmetry of the aforementioned fundamental integrable theories and subsequently connecting them with asymmetrically gauged WZW models, each of which depends on both the gauge fields of the fundamental integrable theories that 
it connects.
The resulting theories are integrable and admit a nice diagrammatic 
representation in terms of webs. The virtue of this diagrammatic 
representation is that given any admissible diagram
one can, at the back of the envelope, directly write down from it the corresponding integrable $\s$-model action (cf. section 2.2 of \cite{Georgiou:2020wwg}).  

The diagrammatic 
representation of these webs is the following. 
To each vertex of a diagram we assigned the matrix $\l_i^{-1}$ of one of the aforementioned fundamental integrable theories.\footnote{This is the case because we will eventually fix the gauge by setting the group element of each of the fundamental integrable theories to the identity, i.e. $\tilde g_i=\mathbb{1}$ for $i=1,2,\dots ,n$.} The vertices are  connected with a number of lines with orientation  which are characterised by the integers  $k_{ij}^{(l_{ij})}$. The orientation is taken to be from the vertex $i$ to the vertex $j$ while the superscript $l_{ij}$ counts the number
of different lines originating from vertex $i$  and arriving at vertex $j$.
Each of these lines is being associated to an asymmetrically gauged WZW model at an arbitrary level $k_{ij}^{(l_{ij})}$. This gauged WZW model is, as usual, a functional of  the corresponding group element $g_{ij}^{(l_{ij})}$.  Gauge invariance of the full action is translated to level conservation at the vertices, namely $\hat k_i=\tilde k_i=k_i$. For more details on the construction of the models, as well as on their graphical representation one may consult section 2 of \cite{Georgiou:2020wwg}. 

Subsequently, we considered two more generic settings. In the first one, we focused on the case in which  the deformation matrix is not diagonal in the space of  the fundamental theories (see section 4.1 of \cite{Georgiou:2020wwg}).
In the second, we examined the case in which,  although the deformation matrix is diagonal in the space of  the fundamental theories,  level conservation at the vertices is relaxed (section 4.2 of \cite{Georgiou:2020wwg}). In both cases the theories at hand are integrable 
only when all the deformation matrices are proportional to the identity in the group space, that is only when all the theories we are coupling are all of the PCM-type.
\begin{figure}[h]
\centering
\includegraphics[scale=1.3]{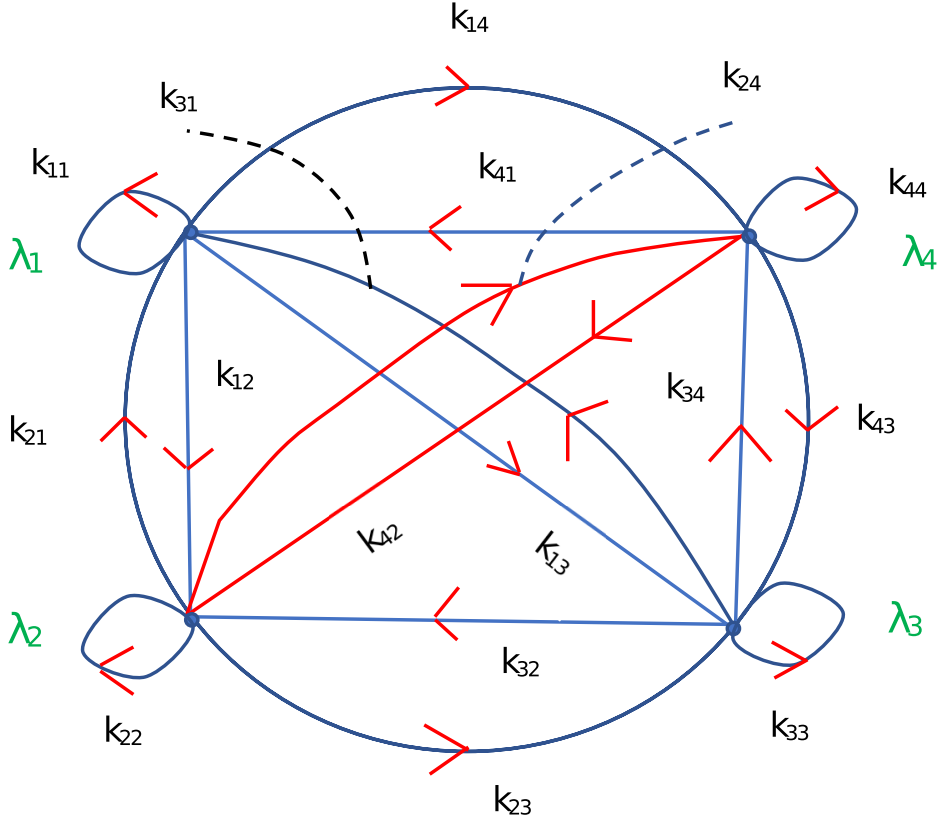}
\caption{{An example of a web representing   an integrable theory built from four of the fundamental  theories, twelve asymmetrically and four vectorially gauged WZW models. The  four fundamental  theories are sitting at the four vertices. 
The lines with orientation (blue lines with red arrows) connecting the vertices are associated with WZW models at arbitrary levels subject to the condition that level conservation at each vertex is imposed.  The red colour for 2 of the lines is just to indicate  that their intersections with the blue lines do not designate vertices. The diagram corresponds to an integrable theory. This very same diagram has been already presented as an example of an integrable web in \cite{Georgiou:2020wwg}}.}
\label{fig:generaltriangle}
\end{figure}

The action after fixing the gauge by setting the group elements of the fundamental theories to unity
and before integrating out the gauge fields reads \cite{Georgiou:2020wwg}
\be\label{gfact}
\begin{split}
&S_{gf}=-{1\ov \pi}\int d^2\s\  A_{+\,a}^{(i)} \,\, (\l^{-1})_{ij}^{ab}\,\, A_{-\,b}^{(j)} +\\
&\sum_{i,j}\sum_{l_{ij}} \Bigg( S_{k_{ij}^{(l_{ij})}}(g_{ij}^{(l_{ij})})
+{k_{ij}^{(l_{ij})}\ov \pi} \int d^2\s \ \Tr \Big(A_-^{(i)} J_{+\,ij}^{(l_{ij})}   - A_+^{(j)} J_{-\,ij}^{(l_{ij})}+ A_-^{(i)} g_{ij}^{(l_{ij})}A_+^{(j)} \big(g_{ij}^{(l_{ij})}\big)^{-1}\Big) \Bigg),
\end{split}
\ee
where the deformation matrix is defined as
\be\label{lambdadef}
(\l^{-1})_{ij}^{ab}={1 \ov 2}(\tilde k_i+\hat k_i)\,\d_{ij}\,\d ^{ab}+E_{ij}^{ab}=\d_{ij}\,(\l_{i}^{-1})^{ab} ,\qquad \,\,\,{\rm since}\,\,\,
E_{ij}^{ab}=\d_{ij} E_i^{ab}.
\ee
In \eqref{gfact} $E_{ij}^{ab}$ denotes the matrix that sits between the currents of the fundamental integrable theories.
Furtermore, the definition of the currents in terms of the group elements is as follows \footnote{Regarding the WZW action and the Polyakov -Wiegmann identity we are following the conventions of \cite{Georgiou:2016urf,Georgiou:2017jfi}.}
\be\label{defj}
J_{+\,ij}^{(l_{ij})}=J_+(g_{ij}^{(l_{ij})})=\partial_+g_{ij}^{(l_{ij})} \big(g_{ij}^{(l_{ij})}\big)^{-1}\, , \qquad J_{-\,ij}^{(l_{ij})}=J_-(g_{ij}^{(l_{ij})})=\big(g_{ij}^{(l_{ij})}\big)^{-1}\partial_-g_{ij}^{(l_{ij})},
\ee

 After integrating out the gauge fields the action becomes
 \be\label{sigma-fin}
\begin{split}
S_{\s-mod.}=-{1\ov \pi}\int d^2\s\    J_+^{(i)} \,\, \,\, \Big( \l^{-1}-{\mathcal D}^T\Big)^{-1}_{ij} \,  J_-^{(j)} +\sum_{i,j}\sum_{l_{ij}}  S_{k_{ij}^{(l_{ij})}}(g_{ij}^{(l_{ij})})\ ,
\end{split}
\ee
where the matrices ${\mathcal D}_{ij}$ and the currents $J_\pm^{(i)}$ are defined as follows
\ba\label{DDT}
&&{\mathcal D}_{ij}=\sum_{l_{ij}}  k_{ij}^{(l_{ij})}  D(g_{ij}^{(l_{ij})}),\qquad \mathcal D^T_{ij}=\sum_{l_{ji}}  k_{ji}^{(l_{ji})} D^T(g_{ji}^{(l_{ji})}),\qquad D^{ab}(g)=\tr{(t^a g \,t^bg^{-1}) }\nonumber\\
&&  J_+^{(i)}=\sum_{ n,l_{in}} k_{in}^{(l_{in})} J_+(g_{in}^{(l_{in})} ),\qq J_-^{(i)}=\sum_{n,l_{ni}} k_{ni}^{(l_{ni})} J_-(g_{ni}^{(l_{ni})} )  \ .
\ea
In \eqref{lambdadef} we have also defined the sum of the WZW level that are directed away from and towards each vertex as 
\be\label{defk}
\tilde k_i=\sum_{j,l_{ij}}k_{ij}^{(l_{ij})}, \qquad \hat k_i=\sum_{j,l_{ji}}k_{ji}^{(l_{ji})}\, , \,\,\,
{\rm where}\,\,\,i=1,\dots,n\,.
\ee
respectively. Furthermore, our conventions for the generators of the group  $G$
 are $\tr (t^a t^b)=\d^{ab}$ and $[t^a,t^b]=i \, f_{abc}t^c$.

\subsection{Integrability}

In \cite{Georgiou:2020wwg} it was argued that for three distinct choices of the deformation matrices $(\l^{-1})_{ij}^{ab}$ the models in \eqref{sigma-fin} are integrable. The first choice is when level conservation at each vertex holds and the matrix $(\l^{-1})_{ij}^{ab}$ is diagonal in the space of models, i.e. $(\l^{-1})_{ij}^{ab}=\d_{ij}(\l_i^{-1})^{ab}$. In the group space the deformation matrix $(\l_i^{-1})^{ab}$
could be either isotropic (the PCM  on the group $G$ or on the symmetric space $G/H$) or non-isotropic (the YB model on group $G$ or on the symmetric space $G/H$). The second choice is when level conservation still holds but the deformation matrix is non-diagonal in the space of models but diagonal and isotropic in the group space (see \eqref{lambda22}). The third choice is when the level conservation is relaxed, that is $\tilde k_i\ne \hat k_i$ and  the deformation matrix is diagonal in the space of models and diagonal and isotropic in the group space, i.e. $(\l^{-1})_{ij}^{ab}=\d_{ij}\d^{ab}\l_i^{-1}$.

As discussed in detail in section 4.2 of \cite{Georgiou:2020wwg}, in all the three cases mentioned above, the equations of motion, $D_-\Big(D_+g_{ij}^{(l_{ij})}\big(g_{ij}^{(l_{ij})}\big)^{-1}\Big)=F_{+-}^{A^{(i)}}$ of the models can be rewritten as follows
\be\label{eomWZW-1aa}
\begin{split}
&\sum_{j, l_{ij}}k_{ij}^{(l_{ij})}D_-\Big( D_+g_{ij}^{(l_{ij})}\big(g_{ij}^{(l_{ij})}\big)^{-1}\Big)=\tilde k_i F_{+-}^{A^{(i)}}\, ,\\
&D_-Y_{+\,\,ij}^{(l_{ij})}=0, \qquad {\rm where}\qquad
Y_{+\,\,ij}^{(l_{ij})}=D_+g_{ij}^{(l_{ij})}\big(g_{ij}^{(l_{ij})}\big)^{-1}-D_+g_{ij}^{(l^{0}_{ij})}\big(g_{ij}^{(l^{0}_{ij})}\big)^{-1} \ .
\end{split}
\ee
Here $g_{ij}^{(l^{0}_{ij})}$ denotes the group element of one of the WZW models that points way from the vertex $i$. This  is used as a reference in the definition of $Y_{+\,\,ij}^{(l_{ij})}$.
Then the first set of equations appearing in the first line of \eqref{eomWZW-1aa}, when combined with the constraints which are obtained from the equations of motion of the gauge fields,
can be recast as zero curvature conditions of certain Lax pairs whose explicit form is different for each of the three cases above and which can be found in  \cite{Georgiou:2020wwg}.
Furthermore, each of the remaining $N-n$ equations of motion in the second line of \eqref{eomWZW-1aa} give an infinite tower of conserved charges. 
\be\label{charges}
Q_{ij}^{(l_{ij})(k)}=\int_0^{2 \pi} \,d\s\, \,\tr\big(Y_{+\,\,ij}^{(l_{ij})}\big)^k,\,\,\,k=1,2,3,\dots ,
\ee
where we have assumes that all fields obey periodic boundary conditions in the interval $\s \in [0,2\pi]$.
As a result, one has as many infinite towers of conserved charges as the number of degrees of freedom of the models, which is the primary definition of a Hamiltonian integrable theory.

However, what is not proved  in \cite{Georgiou:2020wwg} is that the whole set of the conserved charges, those obtained from the monodromy matrices
 that are constructed from the Lax connections and those obtained from the covariantly free  equations of motion
 in the second line of 
\eqref{eomWZW-1aa}, are in involution.
This gap in the proof of integrability is filled in the following sections. In the next section, by employing the Hamiltonian formalism, we show that all the aforementioned conserved charges are, indeed, in involution. Subsequently, in section \ref{3} we show that the complete set of equations of motion \eqref{eomWZW-1aa}, including those in the second line of \eqref{eomWZW-1aa}, can be rewritten as zero curvature conditions of spectral parameter dependent Lax connections.

\section{Hamiltonian formulation  of the webs of integrable theories and charges in involution }\label{2}

In this section, we derive the Hamiltonian densities of the class of  integrable theories presented in \cite{Georgiou:2020wwg}. This class of integrable theories have a simple and transparent diagrammatic representation which allows one, given the relevant diagram,  to immediately write down the 
action of the corresponding integrable $\s$-model.

The action describing these models before integrating out the gauge fields can be found in equations (3.11) and (3.12) of \cite{Georgiou:2020wwg}  and is given in \eqref{gfact}. 
To proceed, we parametrise the group elements $g_{ij}^{(l_{ij})}$  by coordinates $x_{ij}^{(l_{ij})\m}$ , with $\m=1,2,\dots , \dim G$.
Subsequently, we define the matrices
\be
\begin{split}
& L_{ij\, \mu}^{(l_{ij})\, a}  = -i \,{\rm Tr}\Big(t^a \big(g_{ij}^{(l_{ij})}\big)^{-1}\del_\m g_{ij}^{(l_{ij})}\Big)\ ,\qq
R_{ij\, \mu}^{(l_{ij})\, a}  = -i\, {\rm Tr}\Big(t^a \del_\m g_{ij}^{(l_{ij})} \big(g_{ij}^{(l_{ij})}\big)^{-1}\Big) = D_{ij}^{(l_{ij})\,ab}L_{ij\, \mu}^{(l_{ij})\, b} \ 
\end{split}
\ee
where the orthogonal matrix $D$ is defined in \eqref{DDT}.
Taking these definitions into account the action \eqn{gfact} becomes
\ba
&&S_{gf}(x_{ij}^{(l_{ij})}, A^{(i)}_\pm)  =-{1\ov \pi}\int d^2\s\  A_{+\,a}^{(i)} \,\, (\l^{-1})_{ij}^{ab}\,\, A_{-\,b}^{(j)} +
\sum_{i,j}\sum_{l_{ij}} {k_{ij}^{(l_{ij})}\ov 4 \pi}   \int d^2\s \Bigg( {1 \ov 2} R_{ij\, \mu}^{(l_{ij})\, a} R_{ij\, \nu}^{(l_{ij})\, a}  \times\nonumber \\
&&(\dot x_{ij}^{(l_{ij})\m} \dot x_{ij}^{(l_{ij})\n}
- x_{\,\,ij}^{\prime(l_{ij})\m}   x_{\,\,ij}^{\prime(l_{ij})\n}) +
\l_{ij\,\,\m\n}^{(l_{ij})} \dot x_{ij}^{(l_{ij})\m}  x_{\,\,ij}^{\prime(l_{ij})\n}+  2i A_-^{(i)\, a}  R_{ij\, \mu}^{(l_{ij})\, a}(\dot x_{ij}^{(l_{ij})\m}+x_{\,\,ij}^{\prime(l_{ij})\m} )\nonumber\\
&& -2 i A_+^{(j)\, a}  L_{ij\, \mu}^{(l_{ij})\, a}(\dot x_{ij}^{(l_{ij})\m}-x_{\,\,ij}^{\prime(l_{ij})\m} )
 +4 A_+^{(j)\, a} D_{ij}^{(l_{ij})\,ba}A_-^{(i)\, b} \Bigg)\ .
\ea
where $\l_{ij\,\,\m\n}^{(l_{ij})}$  is the antisymmetric couplings of the WZW actions.
From this action one easily deduces the canonical momenta to be
\ba
\label{momenta}
&&\p_{ij\, \mu}^{(l_{ij})}={\partial {\cal L}\ov \partial \dot x_{ij}^{(l_{ij})\m}}=
{k_{ij}^{(l_{ij})}\ov 4\pi} \big( R_{ij\, \mu}^{(l_{ij})\, a} R_{ij\, \nu}^{(l_{ij})\, a}\dot x_{ij}^{(l_{ij})\n} + \l_{ij\,\,\m\n}^{(l_{ij})} x_{\,\,ij}^{\prime(l_{ij})\n}
+ 2 i A_-^{(i)\, a}  R_{ij\, \mu}^{(l_{ij})\, a} - 2 i A_+^{(j)\, a}  L_{ij\, \mu}^{(l_{ij})\, a}\big)\ ,
\nonumber\\
 && P_\pm^{(i)\,a}= {\partial {\cal L}\ov \partial \dot A_\pm^{(i)\,a}}=0 \ .
\ea
In \eqref{momenta} no sum is assumed for the repeated indices.
The above coordinates and momenta obey the usual Poisson bracket relations.
The vanishing of the canonical momenta conjugate to the fields $A_\pm^{(i)\,a}$  imply the existence of constraints
in the Hamiltonian formalism. As a result,  one should invoke Dirac's procedure in order to deal with them. To this end, we derive
the Hamiltonian of our system, in the usual way. It takes the following form
\be
\begin{split}
&  H_0  ={1\ov \pi}\int d\s\  A_{+\,a}^{(i)} \,\, (\l^{-1})_{ij}^{ab}\,\, A_{-\,b}^{(j)} +
\sum_{i,j}\sum_{l_{ij}} {k_{ij}^{(l_{ij})}\ov 4 \pi}   \int d\s \Bigg( {1 \ov 2} R_{ij\, \mu}^{(l_{ij})\, a} R_{ij\, \nu}^{(l_{ij})\, a}  \times \\
&(\dot x_{ij}^{(l_{ij})\m} \dot x_{ij}^{(l_{ij})\n}
+ x_{\,\,ij}^{\prime(l_{ij})\m}   x_{\,\,ij}^{\prime(l_{ij})\n}) -
  2i A_-^{(i)\, a}  R_{ij\, \mu}^{(l_{ij})\, a}x_{\,\,ij}^{\prime(l_{ij})\m}-2 i A_+^{(j)\, a}  L_{ij\, \mu}^{(l_{ij})\, a}
  x_{\,\,ij}^{\prime(l_{ij})\m} 
 -4 A_+^{(j)\, a} D_{ij}^{(l_{ij})\,ba}A_-^{(i)\, b} \Bigg)\ .
\label{H0}
 \end{split}
\ee
Next we define the currents ${\cal J}_{\pm\,ij}^{(l_{ij})\, a} $ to be the same functions of the phase space variables
appearing in the ungauged WZW model \cite{Bowcock}, namely
\be
\label{currents}
\begin{split}
&
-i {\cal J}_{+\,ij}^{(l_{ij})\, a}={1 \ov 2} R_{ij}^{(l_{ij})\, a\, \mu}\left({4\pi \ov k_{ij}^{(l_{ij})}}\p_{ij\, \mu}^{(l_{ij})} - \l_{ij\,\,\m\n}^{(l_{ij})} x_{\,\,ij}^{\prime(l_{ij})\n}\right)+\ha  R_{ij\, \mu}^{(l_{ij})\, a} x_{\,\,ij}^{\prime(l_{ij})\m}
\\
&\qq\qq  
= \ha R_{ij\, \m}^{(l_{ij})\, a}(\dot x_{ij}^{(l_{ij})\m} +x_{\,\,ij}^{\prime(l_{ij})\m} ) + i A_-^{(i)\, a} - i D_{ij}^{(l_{ij})\,ab}A_+^{(j)\, b}\ ,
\\
& i {\cal J}_{-\,ij}^{(l_{ij})\, a}={1 \ov 2} L_{ij}^{(l_{ij})\, a\, \mu}\left({4\pi \ov k_{ij}^{(l_{ij})}}\p_{ij\, \mu}^{(l_{ij})} - \l_{ij\,\,\m\n}^{(l_{ij})} x_{\,\,ij}^{\prime(l_{ij})\n}\right)-\ha  L_{ij\, \mu}^{(l_{ij})\, a} x_{\,\,ij}^{\prime(l_{ij})\m}\\
& \qq\qq
=\ha L_{ij\, \m}^{(l_{ij})\, a}(\dot x_{ij}^{(l_{ij})\m} -x_{\,\,ij}^{\prime(l_{ij})\m} ) - i A_+^{(j)\, a} + i D_{ij}^{(l_{ij})\,ba}A_-^{(i)\, b}\ \  ,
\end{split}
\ee
where $ R_{ij}^{(l_{ij})\, a\, \mu}$ is the inverse of $R_{ij\, \m}^{(l_{ij})\, a}$ and similarly for  $ L_{ij}^{(l_{ij})\, a\, \mu}$.
In addition, we have used the identity $R_{ij\, \m}^{(l_{ij})\, a}R_{ij\, \n}^{(l_{ij})\, a}=L_{ij\, \m}^{(l_{ij})\, a}L_{ij\, \n}^{(l_{ij})\, a}$.

\no
The definitions above ensure that the Poisson brackets (PBs) of the currents take the same form as in the
ungauged WZW theory \cite{Bowcock}. This holds because the PBs of the canonical variables $ x_{ij}^{(l_{ij})\m}$ and $\p_{ij\, \mu}^{(l_{ij})}$ remain unchanged.
After some algebra the Hamiltonian \eqref{H0} can be written as 
\ba\label{-H0}
 && H_0  =- {1\ov 4\pi} \int d\s \   \Bigg( \sum_{i,j}\sum_{l_{ij}} k_{ij}^{(l_{ij})}\Big( {\cal J}_{+\,ij}^{(l_{ij})\, a} {\cal J}_{+\,ij}^{(l_{ij})\, a} + 
{\cal J}_{-\,ij}^{(l_{ij})\, a} {\cal J}_{-\,ij}^{(l_{ij})\, a}\Big)
\nonumber\\
&&+\sum_i (4 \mathbb  J_-^{(i)\, a}A_+^{(i)\, a} +4 \mathbb  J_+^{(i)\, a}A_-^{(i)\, a}+2 \hat k_i A_+^{(i)\, a} A_+^{(i)\, a} +2 \tilde k_i A_-^{(i)\, a} A_-^{(i)\, a} ) -\sum_{ij} 4A_+^{(i)\, a}(\l^{-1})_{ij}^{ab}\,A_-^{(j)\, b}\Bigg)
\nonumber \\
&&
\ea
where we have defined the currents running in and out of any of the vertices as 
\be\label{mathjs}
 \mathbb  J_-^{(i)}=\sum_{n,l_{ni}} k_{ni}^{(l_{ni})}  
{\cal J}_{-\,ni}^{(l_{ni})}\qq{\rm and} \qq\mathbb J_+^{(i)}=\sum_{ n,l_{in}} k_{in}^{(l_{in})} {\cal J}_{+\,in}^{(l_{in})},\,\,\,\,i=1,\dots,n
\ee
 respectively.

Based on the definitions \eqref{currents} the PBs of the currents and the gauge fields are given by 
\be\label{PB-currents}
\begin{split}
&  \{ {\cal J}_{\pm\,ij}^{(l_{ij})\, a}(\s_1),{\cal J}_{\pm\,ij}^{(l_{ij})\, b}(\s_2) \}
= -{2\pi\ov k_{ij}^{(l_{ij})}} \big( i f_{abc}{\cal J}_{\pm\,ij}^{(l_{ij})\, c}(\s_2) \d_{12} \pm  \d_{ab} \d'_{12}\big)\ ,\\
&  \{ {\cal J}_{+\,ij}^{(l_{ij})\, a}(\s_1),{\cal J}_{-\,ij}^{(l_{ij})\, b}(\s_2) \}=0 \ , \qq
 \{ A^{(i)\,a}_{\pm}(\s_1),P^{(j)\,b}_{\pm}(\s_2) \}=\d^{ab} \d^{ij}\d(\s_{12})\ ,
\end{split}
\ee
with all other PBs being zero.

To the Hamiltonian \eqref{-H0} one should add arbitrary linear combinations of the primary constraints which appear in the last line of
\eqref{momenta}.   The resulting Hamiltonian then reads
\be\label{Ht}
H=H_0+\int d \s (c^{(i)\,a}_+ P_+^{(i)\,a} +c^{(i)\,a}_- P_-^{(i)\,a})\ .
\ee
The time evolution of the primary constraints generates secondary constraints as
\be
\begin{split}
&  \{ P_+^{(i)\,a},H \}=  4\big( \hat k_i A_+^{(i)\, a} - (\l^{-1})_{ij}^{ab}\,A_-^{(j)\, b} +\mathbb  J_-^{(i)\, a}\big)\approx 0\ ,
\\ \label{sec1}
&  \{ P_-^{(i)\,a},H \}= 4\big(\tilde k_i A_-^{(i)\, a} - (\l^{-1})_{ji}^{ba}\,A_+^{(j)\, b} +\mathbb  J_+^{(i)\, a}\big)\approx 0\ .
\end{split}
\ee
One can easily verify that the time evolution of the secondary constraints, after adding them to the Hamiltonian as we did with the primary ones, gives no additional constraints. 
We conclude that our theories have $n$ sets of four constraints, primary and secondary, given by 
\be
\begin{split}
&\chi_1^{(i)\,a}=P_+^{(i)\,a}\approx 0\ ,
\\
&\chi_2^{(i)\,a}=P_-^{(i)\,a}\approx 0\ ,
\\
& \chi_3^{(i)\,a}= \hat k_i A_+^{(i)\, a} - (\l^{-1})_{ij}^{ab}\,A_-^{(j)\, b} +\mathbb  J_-^{(i)\, a}\approx 0\ ,
\\ \label{conA}
& \chi_4^{(i)\,a}= \tilde k_i A_-^{(i)\, a} - (\l^{-1})_{ji}^{ba}\,A_+^{(j)\, b} +\mathbb  J_+^{(i)\, a}\approx 0 \ .
\end{split}
 \ee

Notice that all the constraints belonging to different sets, i.e. two constraints that have different values of $i$ have  zero PBs among themselves.
Thus, we obtain
\be
\begin{split}
C^{(i j)}_{\a\b}= \{ \chi^{(i)}_ {\a},\chi_{\b}^{(j)} \} =\d^{ij} C_{\a\b}^{(i)},  \quad \a,\b=1,\dots,4\ .
 \end{split}
\label{Cmatrix}
\ee
The precise form of $C_{\a\b}^{(i)}$ can be found in equation (3.37) of \cite{Georgiou:2016urf} and will not be needed in what follows.
What is important is that $C_{\a\b}^{(i)}$ is invertible with its inverse matrix $(C^{(i)})^{-1}_{\a\b}$ acquiring the form
\be
\begin{split}
& (C^{(i)})_{\a\b}^{-1}= 
  \left(  \begin{array}{cccc}
 (C^{(i)})_{11}^{-1}&   (C^{(i)})_{12}^{-1}&(C^{(i)})_{13}^{-1}&(C^{(i)})_{14}^{-1} \\
   (C^{(i)})_{21}^{-1}&   (C^{(i)})_{22}^{-1}& (C^{(i)})_{23}^{-1}&(C^{(i)})_{24}^{-1}\\
     (C^{(i)})_{31}^{-1}  & (C^{(i)})_{32}^{-1}& 0& 0\\
  (C^{(i)})_{41}^{-1}&(C^{(i)})_{42}^{-1}& 0 & 0\\
  \end{array} \right)\  .
 \label{C1matrix}
 \end{split}
\ee
It is the invertibility of $C^{(i)}_{\a\b}$ which ensures that the constraints \eqref{conA} are second class constraints. Here we should stress that the invertibility of each of the constraint matrices $C^{(i)}_{\a\b}$ is true only when the corresponding fundamental theory is  the PCM or the YB on the group space $G$. If the deformation matrix corresponds to the PCM or the YB on the symmetric coset space 
$G/H$ then the constraint matrix is not invertible since some of the constraints become first class due to the gauge symmetry of the model. This case will be treated separately later on.\\
The above form of $(C^{(i)})_{\a\b}^{-1}$ can now be used to show that the definition of the Dirac brackets
\be
\label{DiracB}
 \{ \zeta, \eta \}_{DB}= \{ \zeta, \eta \}- \{ \zeta, \chi^{(i)}_{\a} \} (C^{(ij)})^{-1}_{\a\b} \{ \chi^{(j)}_{\b}, \eta \} \ ,
\ee
implies that the Dirac brackets of the currents ${\cal J}_{\pm\,ij}^{(l_{ij})}$ are identical to their PBs since the aforementioned currents have nonzero PB only with the constraints $\chi_3$ and $\chi_4$, namely 
\be\label{DB=PB}
\{ {\cal J}_{\pm\,ij}^{(l_{ij})\, a}(\s_1),{\cal J}_{\pm\,ij}^{(l_{ij})\, b}(\s_2) \}_{DB}=\{ {\cal J}_{\pm\,ij}^{(l_{ij})\, a}(\s_1),{\cal J}_{\pm\,ij}^{(l_{ij})\, b}(\s_2) \}\ .
\ee
This is referred in \cite{Hollowood:2014rla} as the protection mechanism.

In passing we note that the time evolution of any dynamical quantity $\zeta$ is now given by $\dot \zeta\approx \{\zeta, H\}_{DB}$.
As mentioned above, due to the structure of the matrix \eqn{Cmatrix} and of its inverse
there will be no mixing between the different sets of constraints. Furthermore, we note that since all the constraints \eqref{conA} are generically
second class 
they may be imposed strongly. This means that one can solve \eqref{conA} for the gauge fields and express them in terms of the currents ${\cal J}_{\pm\,ij}^{(l_{ij})}$ and in particular in terms of their linear combinations  $\mathbb  J_\pm^{(i)}$. It is this strategy 
that we will follow from now on, namely we will impose strongly the second class constraints of the theory. This implies that wherever the gauge fields $A_\pm^{(i)}$ appear they should be thought as functions of the currents.

Given the observations of the last paragraph one can rewrite the Hamiltonian density of the Hamiltonian \eqref{-H0}
\ba\label{-Hfin}
&&  {\cal H}_0  = -{1\ov 4\pi}  \  \Bigg( \sum_{i,j}\sum_{l_{ij}} k_{ij}^{(l_{ij})}\Big( {\cal J}_{+\,ij}^{(l_{ij})\, a} {\cal J}_{+\,ij}^{(l_{ij})\, a} + 
{\cal J}_{-\,ij}^{(l_{ij})\, a} {\cal J}_{-\,ij}^{(l_{ij})\, a}\Big)-\sum_i \big({1\ov \tilde k_i}\mathbb  J_+^{(i)\, a}\mathbb J_+^{(i)\, a}+ {1\ov \hat k_i}\mathbb  J_-^{(i)\, a}\mathbb J_-^{(i)\, a}\big)
\nonumber
\\
&&+\sum_{i,j}\big(A^{(i)\,a}_{+} M^{ab}_{ij}A^{(j)\,b}_{+}+A^{(i)\,a}_{-}N^{ab}_{ij} A^{(j)\,b}_{-}\big) \Bigg).
\ea
To get this last equation we have added and subtracted from  the right hand side of \eqref{-H0} the second sum which  appears in the first line of the equation \eqref{-Hfin}. Then the sum of this term
with the terms in the second line of \eqref{-H0} give the second line of \eqref{-Hfin}.
We have also defined the matrices $M^{ab}_{ij}$ and $N^{ab}_{ij}$ as follows
\be\label{MN}
M^{ab}_{ij}={1 \ov \tilde k_i}\Big( (\l^{-1})_{in}(\l^{-T})_{jn}\Big)^{ab}-\hat k_i \d_{ij}\d^{ab},\qq N^{ab}_{ij}={1 \ov \hat k_i}\Big( (\l^{-T})_{ni}(\l^{-1})_{nj}\Big)^{ab}-\tilde k_i \d_{ij}\d^{ab}\ ,
\ee
 where a sum over the repeated index $n$ is assumed. We have have also sticked to our convention that the transposition refers only to the group indices $a$ and $b$.
As mentioned above, in \eqref{-Hfin} one should think of $A^{(i)\,a}_{\pm}$ to be functions of the currents as obtained by solving  the constraints \eqref{conA}. 

A last comment is in order. Let us mention 
that the Hamiltonian density \eqref{-Hfin}  is the sum of two pieces.
The first piece, which is obtained from \eqref{-Hfin} if we set $(\l^{-1})_{ij}^{ab}={1\ov 2}(\hat k_i+\tilde k_i)\d_{ij}\d^{ab}$  , is the Hamiltonian of an asymmetrically gauge CFT which is anomalous free 
if and only if $\hat k_i =\tilde k_i$. The remaining piece is in fact canonically equivalent to the most general $\l$-deformed model of \cite{Georgiou:2018gpe} in the case when $\hat k_i=\tilde k_i$. We will elaborate on this in the sections to follow. Furthermore, notice that the theories described by  \eqref{-Hfin} are not generically integrable. This happen for specific choices of the deformation matrix 
$(\l^{-1})_{ij}^{ab}$. In what follows, we will focus on three such choices.

\subsection{Integrable models with level conservation}\label{2.1}

In this section , we will focus on the models in which level conservation at each separate vertex is imposed, i.e. $\hat k_i=\tilde k_i=k_i$
.\subsubsection{Deformation matrix diagonal in the space of models}\label{2.1.1}
We first consider the integrable  models studied in detail in  section 3 of \cite{Georgiou:2020wwg}. 
For these models the matrix $\l^{-1}$ is assumed to be diagonal in the space of models, that is $\l^{-1}_{ij}=\d_{ij}\l^{-1}_i$ while in the group
space $\l^{-1}_i$ may be the matrix corresponding to one of the following fundamental integrable models,  the PCM,  the YB model, both based on a group $G$, the isotropic $\s$-model on the symmetric coset space $G/H$ or  the YB model on the symmetric space $G/H$. As discussed in \cite{Georgiou:2020wwg}, gauge invariance of the action before fixing the gauge is equivalent to level conservation at each vertex, that is $\hat k_i=\tilde k_i= k_i$. 

Under the aforementioned restrictions the Hamiltonian density slightly simplifies to 
\ba\label{-Hfin-1st}
&&  {\cal H}_0  = -{1\ov 4\pi}  \  \Bigg( \sum_{i,j}\sum_{l_{ij}} k_{ij}^{(l_{ij})}\Big( {\cal J}_{+\,ij}^{(l_{ij})\, a} {\cal J}_{+\,ij}^{(l_{ij})\, a} + 
{\cal J}_{-\,ij}^{(l_{ij})\, a} {\cal J}_{-\,ij}^{(l_{ij})\, a}\Big)-\sum_i {1\ov  k_i}\big(\mathbb  J_+^{(i)\, a}\mathbb J_+^{(i)\, a}+ \mathbb  J_-^{(i)\, a}\mathbb J_-^{(i)\, a}\big)
\nonumber
\\
&&+\sum_{i}A^{(i)\,a}_{+} \Big({1 \ov  k_i} (\l^{-1})_{i}(\l^{-T})_{i}- k_i \,\mathbb 1 \Big)^{ab}A^{(i)\,b}_{+}+A^{(i)\,a}_{-}\Big({1 \ov  k_i} (\l^{-T})_{i}(\l^{-1})_{i}- k_i \,\mathbb 1 \Big)^{ab} A^{(i)\,b}_{-} \Bigg)\ . \nonumber
\\
\ea
Notice that the Hamiltonian density \eqref{-Hfin-1st} becomes the sum of two parts. The first part, appearing in the first line of \eqref{-Hfin-1st}, corresponds to an {\it anomaly free, asymmetrically} gauged conformal field theory
(CFT) with symmetry group ${\overbrace{G\times G\times ...\times G\times G}^{N} \ov \underbrace{G\times ...\times G}_n}$. 
The role of the terms with the negative levels is to project out the corresponding combination of degrees of freedom.
Another way to see this is by setting $(\l_i^{-1})^{ab}=\d^{ab} k_i$. Then the second line of \eqref{-Hfin-1st} vanishes and the action  (see eq. \eqref{gfact}) from which this Hamiltonian originates is that of an anomaly free, asymmetrically gauged CFT consisting of sum of WZW models \cite{Witten:1991mm}. The precise way of embedding each of the $n$  groups $G$ in the full $\overbrace{G\times G\times ...\times G\times G}^{N}$ symmetry group of the WZW models is dictated by \eqref{gfact}, namely every group of the fundamental integrable models associated with a vertex is embedded from the right to those WZW models arriving at the specific vertex and from the left to those WZW models leaving the vertex. This can be seen from \eqref{gfact} after recalling the transformation properties of the WZW group elements, namely
 $g_{ij}^{(l_{ij})}\rightarrow \L_i^{-1}g_{ij}^{(l_{ij})}\L_j$ \cite{Georgiou:2020wwg}.\\
The second part of the Hamiltonian that appears in the second line of \eqref{-Hfin-1st} can be shown to be canonically equivalent to the sum of $n$ independent $\l$-deformed models. Each of the $\l_i$ matrices belongs to one of the four aforementioned cases, the PCM or the YB model based on $G$, the isotropic $\s$-model on the symmetric coset space $G/H$ or  the YB model on the symmetric space $G/H$.  To see this, notice that from \eqref{PB-currents} one can easily derive the PBs of the linear combinations $\mathbb  J_\pm^{(i)\, a}$, which as discussed above are equal to their Dirac brackets, to obtain 
\be\label{PB-calJ}
\begin{split}
&  \{ {1\ov k_i}\mathbb  J_\pm^{(i)\, a}(\s_1),{1\ov k_i}\mathbb  J_\pm^{(j)\, b}(\s_2) \}
= -{2\pi\ov k_i}\d^{ij} \big( i f_{abc}{1\ov k_i} \mathbb  J_\pm^{(j)\, c}(\s_2) \d_{12} \pm  \d_{ab} \d'_{12}\big)\ ,\\
& \{ \mathbb  J_+^{(i)\, a}(\s_1),\mathbb  J_-^{(j)\, b}(\s_2)\}=0, \qq i,j=1,\dots,n \ .
\end{split}
\ee
These PBs imply that one can define $n$ independent currents, ${\cal J}_\pm^{(i)}(g_i)$, with each of these currents depending on a single group element $g_i$. Those currents are defined in terms of their coordinates and their conjugate momenta as in \eqref{currents}, albeit with $k_i$ in the place of $k_{ij}^{(l_{ij})}$, and as a result they obey the same algebra as the one in \eqref{PB-calJ}.
Then by identifying   ${1\ov k_i}\mathbb  J_\pm^{(i)\, a}$ with ${\cal J}_\pm^{(i)}(g_i)$ , that is 
\be\label{can-transf}
{1\ov k_i} \mathbb  J_\pm^{(i)\, a}={\cal J}_\pm^{(i)\, a}(g_i), \qq i=1,\dots,n 
\ee
it is obvious that one can write the second line of  \eqref{-Hfin-1st} as a sum of $n$ independent single $\l$-deformed models. 
The relations connecting the phase space variables appearing in the second line of  \eqref{-Hfin-1st} and those appearing in the sum of the independent single $\l$-deformed models is precisely \eqref{can-transf}. Actually, one should think of \eqref{can-transf} as follows.\\
One can, in principle,  use \eqref{can-transf} to express the canonical coordinates and momenta  of $n$ of the WZW models in terms of the canonical variables of the remaining $N-n$ WZW models and of the canonical variables of the $n$ newly defined group elements $g_i,\,\, i=1,\dots,n$.    Thus, the canonical transformation in the complete phase space of the theories, whose dimensionality is $2N\,dim(G)$,  consists of the aforementioned $2n\, dim(G)$  relations originating from \eqref{can-transf} with the canonical variables of the remaining $N-n$ WZW models left intact. \\One additional comment is in order. By denoting with $H_1$ the first line of 
\eqref{-Hfin-1st}  it is straightforward to show that 
\be\label{inde}
\{\mathbb  J_\pm^{(i)\, a}(\s_1), H_1 \}=0.
\ee
This is consistent with the fact that the $2n\, dim(G)$ phase space variables appearing in the second line of  \eqref{-Hfin-1st} are effectively projected out from the first line of \eqref{-Hfin-1st}. In other words, it is the gauging which reduces the   dynamical phase space variables appearing in the first line of \eqref{-Hfin-1st} from  $2N\, dim(G)$  to $2(N-n)\, dim(G)$.

The fact that the Hamiltonian of our models becomes the sum of $n$ single $\l$-deformed models plus an asymmetrically gauged model based on WZW models on the  group ${\overbrace{G\times G\times ...\times G\times G}^{N} \ov \underbrace{G\times ...\times G}_n}$ is in accordance with the fact already stressed in \cite{Georgiou:2020wwg}. Namely, that the equations of motion for this class of models split into two sets. The first set consists of $n$ equations which when written in terms of the gauge fields $A_\pm^i$ become similar in form to the equations of motion of $n$ independent single $\l$-deformed models. The second set of equations consists of $N-n$ covariantly free equations which are related to 
the $N-n$ degrees of freedom of the ${\overbrace{G\times G\times ...\times G\times G}^{N} \ov \underbrace{G\times ...\times G}_n}$ asymmetrically gauged model. Finally, let us mention that the sum of $n$ independent $\l$- deformed models is, of course, a special case  of the most general $\l$-deformed models of \cite{Georgiou:2018gpe}.\\
\subsubsection*{\underline{Charges in involution}}
\subsubsection*{\underline{{\it Group case}}}
We now proceed to show that all the conserved charges constructed in  \cite{Georgiou:2020wwg} are in involution among themselves. We will first consider the group case. There are two classes of conserved charges. The first class is the one which includes all the charges derived from the $n$ Lax connections of equations (3.8), (3.9), (3.11) and (3.12) of \cite{Georgiou:2020wwg}.  These Lax connections are, in turn, built from the gauge fields $A_\pm^{(i)}$. One may solve the last two equations in \eqref{conA} for the gauge fields in terms of the currents to obtain 
\be\label{Ainvert}
A_+^{(i)}=M^{-1}_{ij}( \mathbb  J_-^{(j)}+\l_{jn}^{-1}{1 \ov \tilde k_n} \mathbb  J_+^{(n)}),\qq A_-^{(i)}=N^{-1}_{ij}( \mathbb  J_+^{(j)}+\l_{nj}^{-T} {1 \ov \hat k_n}\mathbb  J_-^{(n)}),
\ee
where we have suppressed  the group indices and sum over the repeated indices $j$ and $n$ is implied. The matrices $M$ and $N$ are defined in \eqref{MN}. For the case at hand, $\l^{-1}_{ij}=\d_{ij}\l^{-1}_i$,  equation \eqref{Ainvert} simplifies to 
\be\label{Ainvert-diagon}
\begin{split}
&A_+^{(i)}=\Big({1 \ov  k_i} (\l^{-1})_{i}(\l^{-T})_{i}- k_i \,\mathbb 1 \Big)^{-1}( \mathbb  J_-^{(i)}+\l_{i}^{-1}{1 \ov k_i} \mathbb  J_+^{(i)})=M_i^{-1}( \mathbb  J_-^{(i)}+\l_{i}^{-1}{1 \ov k_i} \mathbb  J_+^{(i)}), \\
&A_-^{(i)}=\Big({1 \ov  k_i} (\l^{-T})_{i}(\l^{-1})_{i}- k_i \,\mathbb 1 \Big)^{-1}( \mathbb  J_+^{(i)}+\l_{i}^{-T} {1 \ov k_i}\mathbb  J_+^{(i)})=
N_i^{-1}( \mathbb  J_+^{(i)}+\l_{i}^{-T} {1 \ov k_i}\mathbb  J_+^{(i)})\ .
\end{split}
\ee
In the last equation there is no sum over the index $i$. 
Now given the PBs \eqref{PB-calJ} and the fact that the Dirac brackets of the gauge fields $A_\pm^{(i)}$ are the same with their PBs, when the gauge fields  are expressed in terms of the currents $ \mathbb  J_\pm^{(i)}$, one immediately concludes that the conserved charges built from different Lax pairs are in involution since
\be\label{independence}
\{ A_\pm^{(i)}(\tau,\s_1),A_\pm^{(j)}(\tau,\s')\}=0 \Longrightarrow
 \{\cL^{(i)}_\pm (\tau,\s_1,\zeta_i),\cL^{(j)}_\pm(\tau,\s_2,\zeta_j)\}
=0\ ,\,\,\,\,\,\, i \neq j \ .
\ee
Furthermore, as shown in \cite{Itsios:2014vfa,Lacroix:2017isl}   the spatial part of the Lax connections for each individual $i=1,\dots,n$ assumes the $r/s$ Maillet form \cite{Maillet:1985ek,Maillet:1985ec}. As a result,
the charges constructed from the monodromy matrix built from  the same Lax pair are also in involution among themselves.

The second class of conserved charges is obtained from the covariantly free equations 
\be\label{eomWZW-2}
\begin{split}
D_-Y_{+\,\,ij}^{(l_{ij})}=0, \,\,\, {\rm where}\,\,\,
Y_{+\,\,ij}^{(l_{ij})}=D_+g_{ij}^{(l_{ij})}\big(g_{ij}^{(l_{ij})}\big)^{-1}-D_+g_{ij}^{(l^{0}_{ij})}\big(g_{ij}^{(l^{0}_{ij})}\big)^{-1}={\cal J}_{+\,ij}^{(l_{ij})}-{\cal J}_{+\,ij}^{(l^{(0)}_{ij})}  .
\end{split}
\ee
The number of these covariantly free equations of motion is $N-n$ since for each vertex one of the group elements $g_{ij}^{(l_{ij}^{0})}$ is used as a reference and as a result the corresponding $Y_{+\,\,ij}^{(l^0_{ij})}=0$.
Each of the covariantly free currents above generates an infinite tower of conserved charges given by
\be\label{charges}
Q_{ij}^{(l_{ij})(k)}=\int_0^{2 \pi} \,d\s\, \,\tr\big(Y_{+\,\,ij}^{(l_{ij})}\big)^k,\,\,\,k=1,2,3,\dots .
\ee
The proof that all the charges in \eqref{charges} are in involution requires three steps.
Firstly, we show that all the conserved charges obtained from the same $Y_{+\,\,ij}^{(l_{ij})}$ commute. Secondly, we show that 
the charges obtained from two $Y$s pointing away from the same vertex also commute. Finally, its is obvious that any two charges obtained from two $Y$s pointing away from  two different vertices are in involution since they depend on currents whose PBs are zero, as can be easily seen from \eqref{eomWZW-2}. \\
We start by making the first step. To simplify  the notation we suppress the subscripts and superscript in $Y_{+\,\,ij}^{(l_{ij})}$ and write $Y$ in the place of $Y_{+\,\,ij}^{(l_{ij})}$. 
We also define the charge densities $Y_m=\tr Y^m$ in terms of which the conserved charges are written as $Q^{(k)}=\int_0^{2 \pi} \,d\s\, \,Y_k$. The algebras we will be considering are the $SO(\cal N)$ and $Sp(2\cal N)$ semi-simple algebras, as well as the non-simple 
$U(\cal N)$ and $GL(\cal N)$ algebras, with the latter being non-compact too. At the end of the section, we will comment on the case of the $SU(\cal N)$ algebra which lacks the property that any given power of an algebra element belongs also to the algebra. 

Then, by recalling that the Dirac brackets of the currents are identical to their PBs, one derives for the   PBs of the charge densities the following expression
\be\label{YmYn}
\begin{split}
\{Y_m(\s_1), Y_n(\s_2)\}=m \,n\, \tr (Y^{m-1}(\s_1)t^a) \, \{Y^a(\s_1), Y^b(\s_2)\}\, \tr (Y^{n-1}(\s_2)t^b).
\end{split}
\ee
Using now \eqref{PB-currents} we obtain 
\be\label{YY}
\begin{split}
& \{Y^a(\s_1), Y^b(\s_2)\} = -2\pi i f_{abc}\Big({1\ov k_{ij}^{(l_{ij})}} {\cal J}_{+\,ij}^{(l_{ij})\, c}(\s_2) + 
 {1\ov k_{ij}^{(l^0_{ij})}} {\cal J}_{+\,ij}^{(l^0_{ij})\, c}(\s_2) \Big)\d_{12}\\
& -2 \pi \Big({1\ov k_{ij}^{(l_{ij})}}+{1\ov k_{ij}^{(l^0_{ij})}}\Big) \d_{ab}\d'_{12}.
\end{split}
\ee
We first treat the terms in the first line of \eqref{YY}. We will see that when inserted in \eqref{YmYn} they give a vanishing result. We focus on the term involving ${\cal J}_{+\,ij}^{(l_{ij})\, c}$. The result for the other term  in the first line of \eqref{YY} will be zero in exactly the  same fashion. We rewrite the relevant term ignoring constants as
\be\label{YY-1st}
\begin{split}
&\tr (Y^{m-1}(\s_1)t^a)\, \tr (Y^{n-1}(\s_2)t^b) \,\tr( t_a\,[t_b,t_c])\,{\cal J}_{+\,ij}^{(l_{ij})\, c}(\s_2) \d_{12}=\\
&\tr(Y^{m-1}(\s_1)[t_b,t_c]) \, \tr (Y^{n-1}(\s_2)t^b)\,{\cal J}_{+\,ij}^{(l_{ij})\, c}(\s_2) \d_{12}= \\
&\tr({\cal J}_{+\,ij}^{(l_{ij})}(\s_2)Y^{m-1}(\s_1)Y^{n-1}(\s_2))\d_{12}
-\tr(Y^{m-1}(\s_1){\cal J}_{+\,ij}^{(l_{ij})}(\s_2)Y^{n-1}(\s_2))\d_{12}=0,
\end{split}
\ee
where we have used twice the completeness relation for the matrices, i. e. $(t^a)_{\a\b}(t^a)_{\g\d}=\d_{\a\d}\d_{\b\g}$, as well as the cyclicity of the trace and, most importantly, the presence of the Dirac delta function in order to set $\s_1=\s_2$ inside the traces.

Thus, we see that only the non-ultralocal term in the second line of \eqref{YY} contributes to the PBs of the charge densities. Plugging this term in \eqref{YmYn} and integrating over $\s_1$ and $\s_2$ to obtain the PBs of the currents we arrive at
\begin{eqnarray}\label{currentPB}
&&\{ Q^{(m)},Q^{(n)} \}=-2 \pi \,m \, n\,  \Big({1\ov k_{ij}^{(l_{ij})}}+{1\ov k_{ij}^{(l^0_{ij})}}\Big) \int_{0}^{2 \pi}\int_{0}^{2 \pi} d\s_1 d\s_2
\tr \big(Y^{m-1}(\s_1) Y^{n-1}(\s_2)\big)\times\nonumber\\ 
&&\times\partial_{\s_1}\d(\s_1-\s_2)
=2 \pi \,m \, n\,  \Big({1\ov k_{ij}^{(l_{ij})}}+{1\ov k_{ij}^{(l^0_{ij})}}\Big)  \int_{0}^{2 \pi}d\s_1\,\tr \big(Y^{n-1}(\s_1)\partial_{\s_1}Y^{m-1}(\s_1) \big)\nonumber\\
&&=2 \pi \,m \, n\, {m-1\ov m+n-2} \Big({1\ov k_{ij}^{(l_{ij})}}+{1\ov k_{ij}^{(l^0_{ij})}}\Big)  \int_{0}^{2 \pi}d\s_1\,
\partial_{\s_1}\tr \big((Y^{m+n-2}(\s_1) \big)=0 \ .
\end{eqnarray}
We have, thus, shown that all the conserved charges obtained from the same $Y_{+\,\,ij}^{(l_{ij})}$ are in involution. The same holds for the charged obtained from two different $Y$s, $Y_{+\,\,ij}^{(l_{ij})}$ and $Y_{+\,\,ij'}^{(l_{ij'})}$ with $j'\neq j$, which are pointing away from the same vertex $i$. The only difference is that in \eqref{YY} and \eqref{currentPB} only the second term in the parenthesis, the one associated with the level $k_{ij}^{(l^0_{ij})}$, is present. Finally, as mentioned above, the charges obtained from  two different $Y$s associated to different vertices, namely $Y_{+\,\,ij}^{(l_{ij})}$ and $Y_{+\,\,i'j'}^{(l_{i'j'})}$ with $j'\neq j$ and  $i'\neq i$, are in involution trivially since the currents 
used in their definition have zero PBs altogether. 

The last step in completing the proof that all the local charges that we have constructed are fully in involution is to show that the PBs of the charges built from the Lax connections are in involution with those built from the covariantly free equations \eqref{eomWZW-2}.
To this end, we evaluate the relevant PBs
\begin{eqnarray}\label{mixedPB}
&&\{A_+^{i\, a'}(\s_1), Y_{+\,\,ij}^{(l_{ij})\, b}(\s_2)\}= {1 \ov k_i}(M_i^{-1}\l_i^{-1})^{\a'\a}\{\mathbb  J_+^{(i)\, a},\, {\cal J}_{+\,ij}^{(l_{ij})\, b}-{\cal J}_{+\,ij}^{(l^{(0)}_{ij})\,b}  \}= \nonumber\\
&&{1 \ov k_i}(M_i^{-1}\l_i^{-1})^{\a'\a}\Big(-2\pi \big( i f_{abc}{\cal J}_{+\,ij}^{(l_{ij})\, c}(\s_2) \d_{12} +  \d_{ab} \d'_{12}\big)
+2 \pi \big( i f_{abc}{\cal J}_{+\,ij}^{(l^0_{ij})\, c}(\s_2) \d_{12} +  \d_{ab} \d'_{12}\big) \Big)\nonumber\\
&&={1 \ov k_i}(M_i^{-1}\l_i^{-1})^{\a'\a}(-2 \pi i) f_{abc}Y_{+\,\,ij}^{(l_{ij})\, c}(\s_2)\d_{12}, 
\end{eqnarray}
where we have used \eqref{Ainvert-diagon} for $A_+^{i}$, the fact that all the PBs between the plus and minus currents are zero, as well as the definition of $Y_{+\,\,ij}^{(l_{ij})}$ in terms of the currents \eqref{eomWZW-2}.
Equation \eqref{mixedPB} directly leads to
\begin{eqnarray}\label{AYm}
&&\{A_+^{i\, a'}(\s_1), Y_m(\s_2)\}=n \{A_+^{i\, a'}(\s_1), Y_{+\,\,ij}^{(l_{ij})\, b}(\s_2)\} (Y_{+\,\,ij}^{(l_{ij})})^{n-1\, b}= \nonumber \\
&&2 \pi n {1 \ov k_i}(M_i^{-1}\l_i^{-1})^{\a'\a}[(Y_{+\,\,ij}^{(l_{ij})})^{n-1},\, Y_{+\,\,ij}^{(l_{ij})}]^a=0.
\end{eqnarray}
In the last equality of \eqref{AYm} we have used that the commutator $[Y_{+\,\,ij}^{(l_{ij})},\, Y_{+\,\,ij}^{(l_{ij})}]^a=0$.
In precisely the same way one can show that 
\be\label{AmYm}
\{A_-^{i\, a'}(\s_1), Y_m(\s_2)\}=0 \ .
\ee
Equations \eqref{AYm} and  \eqref{AmYm} directly imply that 
\be\label{LQPB}
\{ \cL^{(i)}_\pm (\tau,\s_1,\zeta_i),\, Y_m(\s_2)\}=0 \Rightarrow\{ \cL^{(i)}_\pm (\tau,\s_1,\zeta_i),\, Q^{(m)}\}=0, \,\,\,\,i=1,\dots,n
\ee
which means that the charges obtained from the monodromy matrices are in involution with those obtained from the covariantly free equations of motion. This ensures the Hamiltonian integrability of the system.
 
 Finally, let us consider the case where the Lie algebra of the group $G$ is $SU(\cal N)$. In this case the PBs of the currents in \eqref{currentPB} turn out to be different from zero, that is   $\{ Q^{(m)},Q^{(n)} \}\neq0$. However, the PB of the charge  densities $ \{Y^a(\s_1), Y^b(\s_2)\}$ obey up to normalisations precisely the same algebra as the charge densities of the currents of the PCM, namely they obey eq. (4.5) of \cite{Evans:1999mj}. This means that, exactly as was done in \cite{Evans:1999mj}, one can systematically construct  conserved current densities of homogeneous spin that have zero PBs and will, thus, produce commuting charges. In our case, the conserved currents will be given by equation  (4.6) of \cite{Evans:1999mj} but with $Y_m$ in the place of ${\cal J}_m$.
 Finally, let us mention that these commuting charge densities are still in involution with the charges obtained from the monodromy matrices due to the first equation in \eqref{LQPB}.

\subsubsection*{\underline{{\it Symmetric coset $G/H$ case}}}
The same conclusion, namely that all conserved charges are in involution,  holds in 
the case in which one or more of the fundamental integrable models is based on the symmetric space $G/H$.
This case can be addressed by taking the corresponding deformation matrices matrices  $\l_i$ to have the following
form, $(\l_i)_H^{ab}=k_i\,\d^{ab}$ and
$(\l_i)_{G/H}^{\a\b}=(\l_i)_{G/H}^{\a\b}$ with all other components zero,
where  the Latin (Greek) indices
take values in the subgroup $H$ (coset $G/H$).\footnote{ For the case of the PCM on the symmetric space the matrix is $(\l_i)_H^{ab}=k_i\d^{ab}$ and $(\l_i)_{G/H}^{\a\b}=(\l_i)_{G/H} \d^{\a\b}$.}
The Hamiltonian analysis of this case remains the same up to equation \eqref{Cmatrix}
with the only difference that one should, in this case, distinguish
the indices of the constraint matrix belonging to the subgroup $H$ from those belonging
to the coset $G/H$. Nevertheless, the matrix $C\sim \d^{ij}$ remains diagonal in the space of models
implying that the gauge fields $A^{{i}}_{\pm}$ and $A^{{j}}_{\pm}$ for $i\ne j$,
from which the  different Lax pairs are built, still commute.
In what follows, we heavily rely on \cite{Bowcock,Hollowood:2014rla}.
The idea is to project the constraints  \eqref{conA} on the subgroup $H$ and coset $G/H$ spaces.
The constraints on the coset space can be used to solve for the gauge fields $A^{{(i)}\, g/h}_{\pm}$ in terms of the currents $ \mathbb  J_-^{(i)\, g/h}$, exactly as we did in the group case of the previous section. However, the projection on the subgroup
results to constraints identical to the case of gauged WZW model \cite{Bowcock,Hollowood:2014rla}, namely
\be
\begin{split}
&\hat \chi_1^{(i)\,h}=P_\tau^{(i)\,h}\approx 0\ ,
\\
&\hat \chi_2^{(i)\,h}=P_\s^{(i)\,h}\approx 0\ ,
\\
&\hat  \chi_3^{(i)\,h}=  2 k_i A_\s^{(i)\, h} +\mathbb  J_-^{(i)\, h}\approx 0\ ,
\\ \label{conB}
& \hat \chi_4^{(i)\,h}= -2 k_i A_\s^{(i)\, h} +\mathbb  J_+^{(i)\, h}\approx 0 \ .
\end{split}
 \ee
The matrix of the PBs of the constraints above has zero determinant which is in accordance with the fact that the coset theory has a
local gauge symmetry with symmetry group $H$.
As pointed out in \cite{Bowcock,Hollowood:2014rla}, two out of the four constraints, namely $P_\s^{(i)\,h}\approx 0$ and $2 k_i A_\s^{(i)\, h} +\mathbb  J_-^{(i)\, h}\approx 0$ can be chosen to be second class and as a result can be imposed strongly on the phase space to
eliminate $P_\s^{(i)\,h}$ and $A_\s^{(i)\, h}$ from the physical space. Consequently, one is left with the following two first class constraints $P_\tau^{(i)\,h}\approx 0$ and $\mathbb  J_+^{(i)\, h}+\mathbb  J_-^{(i)\, h}\approx 0$
 which reflect the invariance of the action with respect to the gauge symmetry  group $H$.
 What is important for our purposes is that the protection mechanism is still at work preventing also the PBs of the projection of the currents in the subgroup $H$, that is   $\mathbb  J_\pm^{(i)\, h}$ from being modified.
 
 Due to the fact that  the PBs of {\it all} the components of  the currents  $\mathbb  J_\pm^{(i)\, a}$ remain unchanged, the proof that all the conserved charges are in involution is similar to that of the group case presented in  the previous section. By imposing strongly the constraints in the way mentioned in the previous paragraph the analogue of \eqref{independence} straightforwardly becomes
 \be\label{independence-coset}
\{ A_\s^{(i)}(\tau,\s_1),A_\s^{(j)}(\tau,\s')\}=0 \Longrightarrow
 \{\cL^{(i)}_\s (\tau,\s_1,\zeta_i),\cL^{(j)}_\s(\tau,\s_2,\zeta_j)\}
=0\ ,\,\,\,\,\,\, i \neq j \ .
\ee
 which implies that the monodromy matrices obtained from the two spatial parts of the Lax connections are in involution. To obtain the last equation we have used the explicit form of the Lax connections given in equations (3.11) and (3.12) of \cite{Georgiou:2020wwg}.
Furthermore, the part of the proof for the charges obtained from the covariantly free equations of motion remains exactly the same as in the group case. Finally,  one can perform the same steps which led from \eqref{mixedPB} to \eqref{AmYm}. The only difference is that one 
has $A_\s^{(i)}(\tau,\s_1)$ in the place of $A_\pm^{i\, a'}(\s_1)$.\footnote{The reason for considering only the 
$\s$ component of the gauge field $A_\pm^{i\, a'}(\s_1)$ and not both of its components is that one should eventually gauge fix the gauge symmetry by choosing $A_\tau^{(i)\,h}(\tau,\s_1)\approx 0$. Notice that the 
$\s$ component  of the gauge field is enough since it is the one which enters the construction of the monodromy matrix.} Then by considering the subgroup and coset projections of $A_\s^{(i)}(\tau,\s_1)$  separately, one straightforwardly derives
\be\label{LQPB-coset}
\{ \cL^{(i)}_\s (\tau,\s_1,\zeta_i),\, Y_m(\s_2)\}=0 \Rightarrow\{ \cL^{(i)}_\s (\tau,\s_1,\zeta_i),\, Q^{(m)}\}=0, 
\ee
where, as usual, only should think of the gauge fields as functions of the currents. 
This completes the proof that the whole set of conserved charges are also in involution in the case where one or more of the fundamental theories living at the vertices are based on the symmetric coset space $G/H$.

\subsubsection{Deformation matrix non-diagonal in the space of models}\label{2.1.2}

In this section, we consider the more general case in which the deformation matrix takes the following non-diagonal form in the space of models while it remains diagonal in the group space
\be
\label{lambda22}
\begin{split}
\tilde \l^{-1}_{ij} =
\left(       \begin{array}{cccc}
            \l^{-1}_{11} & 0 &  \cdots  &0  \\
             \l^{-1}_{21} & 0 &\cdots  & 0\\
              \vdots  & \vdots  & \ddots & \vdots \\
              \l^{-1}_{(n-1)1}  & 0 & \cdots &  0\\
               0 & \l^{-1}_{n2} & \cdots  & \l^{-1}_{nn}\\
           \end{array}
         \right)\  ,
\end{split}
\ee
where we have redefined 
\be\label{redefA}
\tilde A_-^{(i)}=\sqrt{\tilde k_i} A_-^{(i)},\qq \tilde A_+^{(i)}=\sqrt{\hat k_i} A_+^{(i)},\qq \tilde \l^{-1} _{ij}={1\ov \sqrt{\tilde k_i \hat k_i}}\l^{-1} _{ij}\ .
\ee
Subsequently one should  set $\tilde k_i=\hat k_i=k_i$.
To prove that the theory is integrable, one should rely on \cite{Georgiou:2018gpe}. The Lax connections equivalent to  the equations of motion in the first line of \eqref{eomWZW-1aa} are given by 
\be
\begin{split}
\label{Lax1}
&\cL^{(1)}_+=\sum_{i=1}^{n-1}c_+^{(i)}(z) \tilde  A_+^{(i)} ,\qquad
\cL^{(1)}_-=z \tilde A_-^{(1)} \\
&\cL^{(2)}_-=\sum_{i=2}^{n}c_-^{(i)}(z) \tilde  A_-^{(i)} ,\qquad
\cL^{(2)}_+=z \tilde A_+^{(n)} 
\end{split}
\ee
where the expressions for  $c_+^{(i)}(z)$ and $c_-^{(i)}(z)$ can be found in section 4.1 of  \cite{Georgiou:2020wwg}.
As in the models of the previous section, the remaining covariantly free equations of motion generate infinite towers of conserved charges. 
The aim of this section is to prove that all the charges, those built from the Lax pairs and those built from the covariantly free equations of motion,
are in involution, ensuring thus integrability of the theories at hand. 
The proof that the charges obtained from the covariantly free equations of motion \eqref{eomWZW-2} are in involution among themselves is unaltered compared to the models of the previous section, see \eqref{currentPB}. Furthermore, the charges obtained from the Lax connections of \eqref{Lax1} are also in involution among themselves. This has been proved in \cite{Georgiou:2019plp} where it was shown that the PBs of the spatial part of the Lax pairs acquire the $r/s$ Maillet form. 

It remains to show that the charges built from the Lax connections are in involution with those in \eqref{charges} which are obtained from the covariantly free equations of motion. To this end, we use the more general expression for the gauge fields in terms of the currents presented in \eqref{Ainvert}. From this equation we obtain
\begin{eqnarray}\label{mixedPB-2}
&&\{A_+^{i\, a'}(\s_1), Y_{+\,\,ij}^{(l_{ij})\, b}(\s_2)\}= {1 \ov k_n}(M_{ij}^{-1}\l_{jn}^{-1})^{\a'\a}\{\mathbb  J_+^{(n)\, a},\, {\cal J}_{+\,ij}^{(l_{ij})}-{\cal J}_{+\,ij}^{(l^{(0)}_{ij})}  \}= \nonumber\\
&&{1 \ov k_i}(M_{ij}^{-1}\l_{ji}^{-1})^{\a'\a}\Big(-2\pi \big( i f_{abc}{\cal J}_{+\,ij}^{(l_{ij})\, c}(\s_2) \d_{12} +  \d_{ab} \d'_{12}\big)
+2 \pi \big( i f_{abc}{\cal J}_{+\,ij}^{(l^0_{ij})\, c}(\s_2) \d_{12} +  \d_{ab} \d'_{12}\big) \Big)\nonumber\\
&&={1 \ov k_i}(M_{ij}^{-1}\l_{ji}^{-1})^{\a'\a}(-2 \pi i) f_{abc}Y_{+\,\,ij}^{(l_{ij})\, c}(\s_2)\d_{12}, 
\end{eqnarray}
where in order to pass from the first to the second line of \eqref{mixedPB-2} we have used the fact that the only non-zero PB is when $n=i$. This can be straightforwardly seen from \eqref{mathjs}.
Then the argument proceeds unchanged from the analogue of \eqref{mixedPB} to \eqref{LQPB} resulting to 
\be\label{LQPB-2}
\{ \cL^{(1,2)}_\pm (\tau,\s_1,\zeta_i),\, Q^{(m)}\}=0, \,\,\,\, 
\ee
where now $ \cL^{(1,2)}_\pm$ are those appearing in \eqref{Lax1}.
This step completes the proof that also in the case where the deformation matrix is given by \eqref{lambda22} the whole set of conserved charges are in involution among themselves.

Let us close this section by making a last comment. Note that the Hamiltonian of the theories considered in this section is given by \eqref{-Hfin} after imposing the level conservation condition $\tilde k_i=\hat k_i=k_i$.\\
Again, the aforementioned  Hamiltonian is canonically equivalent
to the sum 
of the Hamiltonian of an ${\overbrace{G\times G\times ...\times G\times G}^{N} \ov \underbrace{G\times ...\times G}_n}$  {\it anomaly free, asymmetrically}  gauged CFT ($1^{st}$ line of \eqref{-Hfin})
and of the one 
describing the {\it integrable} models constructed in \cite{Georgiou:2018gpe} ($2^{nd}$ line of \eqref{-Hfin}).
\subsection{Integrable models with level conservation relaxed}\label{2.2}

In this section, we focus on the integrable models presented in section 4.2 of  \cite{Georgiou:2020wwg}. In the aforementioned work one can find the expressions for the flat Lax connections which encode part of the equations of motion, as well as the charges derived from the remaining equations of motion which take the form of covariantly free quantities.
The main feature of these models is that we drop the requirement that the sum of the levels of the asymmetrically WZW models pointing towards any of the vertices is equal to the sum of the levels  pointing away from it. The integrability of these models is guaranteed if 
the deformation matrix is diagonal in both the space of models and in the group space, that is $(\l^{-1})_{ij}^{ab}= \l_i^{-1}\d_{ij}\d^{ab}$.
In this case the Hamiltonian becomes
\ba\label{-Ham3}
&&  {\cal H}_0  =- {1\ov 4\pi}  \  \Bigg( \sum_{i,j}\sum_{l_{ij}} k_{ij}^{(l_{ij})}\Big( {\cal J}_{+\,ij}^{(l_{ij})\, a} {\cal J}_{+\,ij}^{(l_{ij})\, a} + 
{\cal J}_{-\,ij}^{(l_{ij})\, a} {\cal J}_{-\,ij}^{(l_{ij})\, a}\Big)-\sum_i \big({1\ov \tilde k_i}\mathbb  J_+^{(i)\, a}\mathbb J_+^{(i)\, a}+ {1\ov \hat k_i}\mathbb  J_-^{(i)\, a}\mathbb J_-^{(i)\, a}\big)
\nonumber
\\
&&+\sum_{i}A^{(i)\,a}_{+} \Big({1 \ov  \tilde k_i} (\l^{-1}_{i})^2- \hat k_i \,\mathbb 1 \Big)A^{(i)\,a}_{+}+A^{(i)\,a}_{-}\Big({1 \ov  \hat k_i} (\l^{-1}_{i})^2- \tilde k_i \,\mathbb 1 \Big) A^{(i)\,a}_{-} \Bigg)\ . 
\ea
Furthermore, the PBs of  the currents $\mathbb J_\pm^{(i)}$ are slightly modified compared to \eqref{PB-calJ} and read
\be\label{PB-calJ-3}
\begin{split}
&  \{ {1\ov \tilde k_i}\mathbb  J_+^{(i)\, a}(\s_1),{1\ov  \tilde  k_i}\mathbb  J_+^{(j)\, b}(\s_2) \}
= -{2\pi\ov \tilde k_i}\d^{ij} \big( i f_{abc}{1\ov \tilde k_i} \mathbb  J_+^{(j)\, c}(\s_2) \d_{12} +  \d_{ab} \d'_{12}\big)\ ,\\
& \{ {1\ov \hat k_i} \mathbb  J_-^{(i)\, a}(\s_1),{1\ov \hat k_i} \mathbb  J_-^{(j)\, b}(\s_2) \}
= -{2\pi \ov \hat k_i}\d^{ij} \big( i f_{abc}{1\ov \hat k_i} \mathbb  J_-^{(j)\, c}(\s_2) \d_{12} -  \d_{ab} \d'_{12}\big)\ ,\\
& \{ \mathbb  J_+^{(i)\, a}(\s_1),\mathbb  J_-^{(j)\, b}(\s_2)\}=0 \ .
\end{split}
\ee

As in the previous sections, one can show that all the conserved charges, those built from the covariantly free equations of motion as well as those obtained from the $n$ independent Lax  pairs, which can be found in section 4.2 of  \cite{Georgiou:2020wwg}, are fully in involution. The proof follows closely the steps of section \ref{2.1.1} for the case of diagonal 
in the space of models deformation matrices. The only difference here is that in \eqref{Ainvert-diagon}, \eqref{mixedPB} and \eqref{AYm} one should keep track of the fact that  the levels $\hat k_i$ and $\tilde k_i$ are different. However, it is easily seen that this small complication does not change the essence of the proof, namely  that all charges commute.

To summarise, we have seen that in all the three cases of integrable models presented in  \cite{Georgiou:2020wwg} the conserved charges are in involution ensuring thus that the corresponding theories are integrable. In the next section, we will also see that in fact the {\it full set} of the equations of motion in all the aforementioned three cases can be recast as zero curvature conditions of certain Lax connections. This is, of course, in complete agreement with the results of the present section and settles the question of integrability of our theories even more definitely.

\section{Lax pairs for all the equations of motion}\label{3}

In this section, we show that the full set of equations of motions of our models can be recast as zero curvature conditions.
This fact is in complete agreement with the results of the previous section where it was proven that there exist as many infinite towers of conserved charges as the degrees of freedom of the theory, which moreover are in involution. As discussed in section 3.2 of \cite{Georgiou:2020wwg} the equations of motion, $D_-\Big(D_+g_{ij}^{(l_{ij})}\big(g_{ij}^{(l_{ij})}\big)^{-1}\Big)=F_{+-}^{A^{(i)}}$, can be rewritten as follows
\be\label{eomWZW-1a}
\begin{split}
&\sum_{j, l_{ij}}k_{ij}^{(l_{ij})}D_-\Big( D_+g_{ij}^{(l_{ij})}\big(g_{ij}^{(l_{ij})}\big)^{-1}\Big)=\tilde k_i F_{+-}^{A^{(i)}}\, ,\\
&D_-Y_{+\,\,ij}^{(l_{ij})}=0, \qquad {\rm where}\qquad
Y_{+\,\,ij}^{(l_{ij})}=D_+g_{ij}^{(l_{ij})}\big(g_{ij}^{(l_{ij})}\big)^{-1}-D_+g_{ij}^{(l^{0}_{ij})}\big(g_{ij}^{(l^{0}_{ij})}\big)^{-1} \ .
\end{split}
\ee
In the same work, it was shown that for certain choices of the deformation matrix $(\l^{-1})_{ij}^{ab}$-those addressed in sections \ref{2.1.2}, \ref{2.1.2} and \ref{2.2} of the present work-
the equations in the first line of \eqref{eomWZW-1a} imply the existence of spectral parameter dependent Lax connections.
The corresponding Lax connections are given in sections 3.1, 4.1 and 4.2 of \cite{Georgiou:2020wwg} for the three integrable cases mentioned above.

However, this was not achieved for the remaining equations of motion, those in the second line of \eqref{eomWZW-1a}.
Here we see that this is indeed possible. For each of the $N-n$ different $Y_{+\,\,ij}^{(l_{ij})}$ one may define the following Lax pair
\be\label{Lax-rem}
\cL_{+\,\,ij}^{(l_{ij})}=D_+g_{ij}^{(l_{ij})}\big(g_{ij}^{(l_{ij})}\big)^{-1}+A_+^{(i)}+{Y_{+\,\,ij}^{(l_{ij})} \ov \zeta_{ij}^{(l_{ij})}},\qquad  \cL_{-\,\,ij}^{(l_{ij})}= A_-^{(i)} \ ,\qq \forall\,\,\,\, l_{ij}\ne l_{ij}^0
\ee
where $\zeta_{ij}^{(l_{ij})}$ is the spectral parameter. One should remember that at each vertex one of the group elements, namely $g_{ij}^{(l_{ij}^0)}$ is used as a reference and should be exempted from \eqref{Lax-rem} since in this case $Y_{+\,\,ij}^{(l_{ij}^0)}=0$. 
This fits with the fact that \eqref{Lax-rem} supply Lax connections for only the $N-n$ degrees of freedom. \\
Then the zero curvature condition $\del_+  \cL_{-\,\,ij}^{(l_{ij})}-\del_-  \cL_{+\,\,ij}^{(l_{ij})}-[\cL_{+\,\,ij}^{(l_{ij})},\cL_{-\,\,ij}^{(l_{ij})}]=0$ is equivalent to
\ba\label{Lax-rem-2}
&&\del_+  \cL_{-\,\,ij}^{(l_{ij})}-\del_-  \cL_{+\,\,ij}^{(l_{ij})}-[\cL_{+\,\,ij}^{(l_{ij})},\cL_{-\,\,ij}^{(l_{ij})}]=0 \Longleftrightarrow\nonumber
\\
&&F_{+-}^{(i)}-D_-\Big(D_+g_{ij}^{(l_{ij})}\big(g_{ij}^{(l_{ij})}\big)^{-1}\Big)- {D_-Y_{+\,\,ij}^{(l_{ij})} \ov  \zeta_{ij}^{(l_{ij})}}=0 \Longleftrightarrow\nonumber\\
&&D_-\Big(D_+g_{ij}^{(l_{ij})}\big(g_{ij}^{(l_{ij})}\big)^{-1}\Big)=F_{+-}^{(i)}\,\,\,\,\, {\rm and } \,\,\,\,\,D_-Y_{+\,\,ij}^{(l_{ij})}=0 \ .
\ea
Consequently, the full set of equations of motion can be recast as zero curvature conditions. The Lax pairs associated with the first line of \eqref{eomWZW-1a} can be found in sections 3.1, 4.1 and 4.2 of \cite{Georgiou:2020wwg} for the three cases of integrable deformation matrices, while those associated with the second line of \eqref{eomWZW-1a} are given by \eqref{Lax-rem}.
To complete the proof of integrability in this context, one should prove that the charges derived from the Lax pairs  above,  or certain combinations of these charges, are in involution. Here we refrain from doing this since a similar task has already been undertaken in the previous sections for the two sets of conserved charges resulting from \eqref{eomWZW-1a}.
Nevertheless, it would clearly be interesting to explore the implications of the existence of the Lax connection of \eqref{Lax-rem} in more detail, and in particular the relevance it might have to the gauged WZW models.




\section{Conclusions}\label{4}

In this work we have presented the Hamiltonian formulation of the recently constructed integrable models of \cite{Georgiou:2020wwg}. For all the cases with level conservation, the Hamiltonian turns out to be canonically equivalent to the Hamiltonian
of an asymmetrically gauged 
${\overbrace{G\times G\times ...\times G\times G}^{N} \ov \underbrace{G\times ...\times G}_n}$ CFT plus 
the Hamiltonian of  the most general  $\lambda$-deformed model based on the group $\underbrace{G\times ...\times G}_n$ or on $\underbrace{G\times ...\times G}_{n_1+n_2}\times\underbrace{G/H\times ...\times G/H}_{n_3+n_4} $ ($n=n_1+n_2+n_3+n_4$) when coset spaces are also present. Using the Hamiltonian formalism, we have proved that  for the three integrable cases addressed in sections \ref{2.1.1}, \ref{2.1.2} and \ref{2.2} of the present work, the full set of conserved charges of these models are in involution. Finally, we have shown that the complete set of equations of motion of the theories can be put in the form of zero curvature Lax connections. 
These results settle in an affirmative way the question regarding the  integrability of the models presented in \cite{Georgiou:2020wwg} 
and clarify the  relation of our models to the ones already present in the literature.

There are several directions for future study. One may, for example, calculate the 2 and 3-point correlation functions involving currents and primary operators as exact functions of the deformation parameters for the class of integrable models analysed in this work. Furthermore, it would be interesting to understand  which  open string boundary conditions preserve integrability. For the single $\l$-deformed model this was achieved in \cite{Driezen:2018glg} while for the $\lambda$-deformed models based on the group $G\times G$
and for both equal  \cite{Georgiou:2016urf} and unequal levels \cite{Georgiou:2017jfi} the corresponding analysis was done in \cite{Pope}.
In the case of our models, we expect the existence a big number of different D-brane configurations that preserve the bulk integrability. 
Finally, one might want to consider our theories on the upper half plane as was done recently in  \cite{Sfetsos:2021pcs} for the single $\l$-deformed model.

\section*{Acknowledgments}
We would like to thank  P. Panopoulos, G. P. D. Pappas, K. Sfetsos and especially K. Siampos  for useful discussions. 
This project has received funding from the Hellenic Foundation for Research and Innovation
(H.F.R.I.) and the General Secretariat for Research and Innovation (G.S.R.I.), under grant
agreement No 15425. 






\end{document}